\documentclass[11pt]{article}
\usepackage{hyperref,graphicx,amsmath}
\pdfoutput=1

\newcommand\eg{e.g.\ }
\newcommand\ie{i.e.\ }
\newcommand\Pra{\ensuremath{\textit{Pr}}}  
\newcommand\Rey{\ensuremath{\textit{Re}}}  
\newcommand\Ray{\ensuremath{\textit{Ra}}}  
\newcommand\BEQN{\begin{equation*}}
\newcommand\EEQN{\end{equation*}}

\begin{document}
\title{Liquid oil painting: Free and forced convection in an enclosure with mechanical and thermal forcing}
\author{Gregory J. Sheard\hspace{2cm} Kean Y. Wong\\
Department of Mechanical and Aerospace Engineering,\\
Monash University, VIC 3800, Australia
\and Martin P. King\\
Bjerknes Centre for Climate Research,\\
Uni Research, NO~5007, Bergen, Norway}

\maketitle


\begin{abstract}
A fluid dynamics video is linked to this article, which have been submitted to the Gallery of Fluid Motion as part of the $65$th American Physical Society meeting of the Division of Fluid Dynamics, held in San Diego, California, USA, over $17$-$20$ November $2012$. The video serves to visualize flows generated in a rectangular enclosure that are subjected to both mechanical and thermal forcing through a common horizontal boundary. This system exhibits features consistent with either horizontal convection or lid-driven cavity flows depending on the ratio between thermal and mechanical stirring, and three different cases are visualized in the linked videos.
\end{abstract}

\section{Introduction and video links}
The video contained in this submission depicts a combination of thermal forcing through a horizontal convection mechanism \cite{MulGrifHugh-2004-jfm, HughGriff-2008-arfm} and mechanical forcing through a lid-driven mechanism \cite{ShankarDeshpande-arfm2000-driven-cavity} being applied to a quiescent body of fluid contained within a rectangular enclosure. In the video, both thermal and mechanical forcing are applied through the bottom boundary. Hence the depicted flow configuration combines both horizontal convection dynamics and lid-driven cavity dynamics, and may serve as a model for global overturning ocean current systems such as the Meridional Overturning Circulation (MOC) \cite{TailleuxRouleau-TellusA-2010}. Both a
\href{SheardWongKing_2012_APS_DFD_GFM_74553_high.mp4}{high-resolution version of the video (720p, 19.0~MB)} and a
\href{SheardWongKing_2012_APS_DFD_GFM_74553_small.mp4}{lower-resolution version (6.97~MB)} are available.

\section{Methodology}
The flow sequences are obtained numerically using an incompressible spectral-element solver combined with a Boussinesq treatment of buoyancy \cite{SheardKing2011amm-horiz-conv}. The results are also representative of configurations where forcing is imposed along the top boundary, with thermal forcing being inverted (because cooler fluid descends with the same dynamics as hotter fluid rises).

The flows are visualized by plotting contours of the temperature field, which are proportional to buoyancy and inversely proportional to density in the Boussinesq approximation. Variations in the temperature field are emphasized by slightly elevating the visualization plane in the out-of-plane direction according to temperature, and lighting the surface elevation to produce a three-dimensional effect.

The enclosure has a ratio of height to width of $0.16$, and the width $L$ serves as the reference length for definitions of both the Rayleigh number and Reynolds number. The Rayleigh number characterizes the strength of the thermal forcing applied as a linearly varying temperature distribution along the bottom boundary, and is defined as
\BEQN
\Ray = \frac{g\alpha\,\delta T\,L^3}{\nu\kappa_T},
\EEQN
where $g$ is the gravitational acceleration, $\delta T$ is the temperature difference imposed along the forcing boundary, $\nu$ is the fluid kinematic viscosity, $\alpha$ is the fluid volumetric expansion coefficient, and $\kappa_T$ is the fluid thermal diffusivity. The Prandtl number of the fluid is defined as
\BEQN
\Pra = \frac{\nu}{\kappa_T},
\EEQN
and is fixed at $\Pra=6.14$ for all three cases in this video, which corresponds to water at standard laboratory conditions.

The Reynolds number characterizes the strength of mechanical stirring applied to the fluid through a constant horizontal motion of the bottom boundary from left to right (in the direction of increasing heating, and therefore increasing added buoyancy). It is defined as
\BEQN
\Rey = \frac{VL}{\nu},
\EEQN
where $V$ is the velocity of the moving boundary.

\section{Video contents}
Three cases are depicted. These sequences, which each run for approximately $40$ seconds, show the transient startup phase of the flow under different conditions as follows:
\begin{enumerate}
  \item $\Ray=10^{10}$, $\Rey=4\,072$,
  \item $\Ray=10^{10}$, $\Rey=5\,700$,
  \item $\Ray=10^{10}$, $\Rey=7\,329$.
\end{enumerate}

The common high Rayleigh number ensures that in the absence of any mechanical stirring (\ie $\Rey=0$) the flow is in a convection-dominated regime. At this condition the long-timescale equilibrium state features continuous time-dependent eruption of buoyant plumes of hot fluid at the heated right-hand end of the enclosure, driving a counter-clockwise overturning circulation in the enclosure that is completed by a diffusive return of fluid at the cooler left end of the enclosure. The three Reynolds numbers straddle a zone in which a gradual shift occurs from a flow with overturning circulation dominated by natural convection (Case~1), through a mixed regime (Case~2) to a flow primarily exhibiting features associated with forced convection and mechanical base-driven stirring (Case~3).
\begin{figure}
\centering
\begin{tabular}{l}
  (\textit{a}) Case~1: $\Ray=10^{10}$, $\Rey=4\,072$\\[3pt]
  \includegraphics[width=0.8\columnwidth]{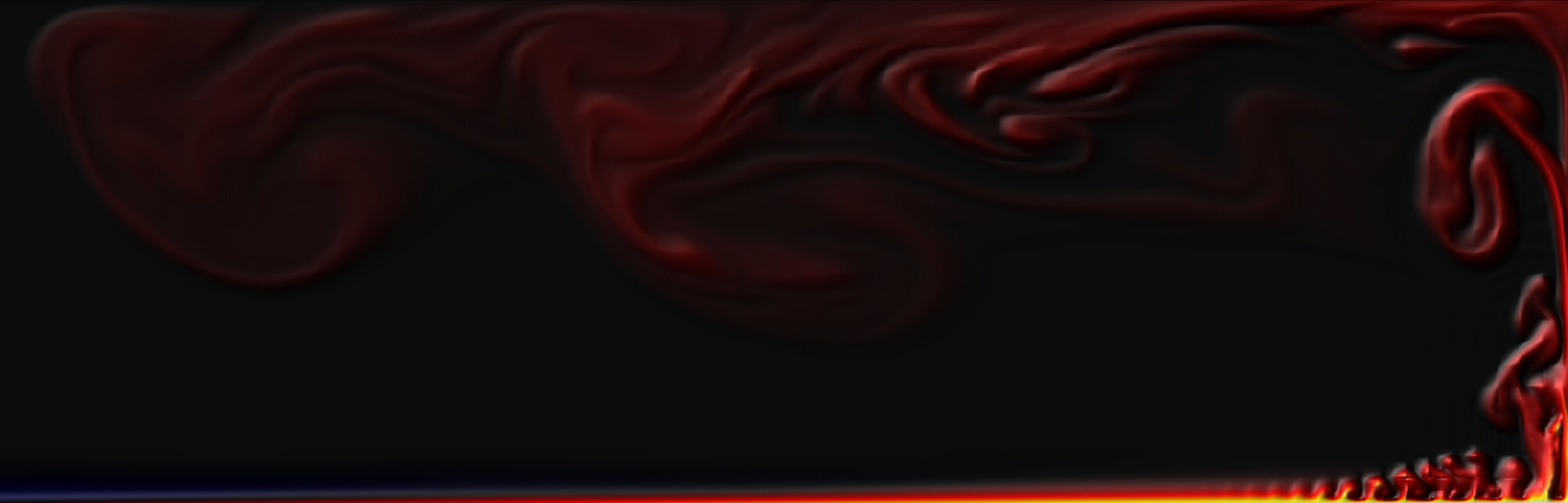} \\[6pt]
  (\textit{b}) Case~2: $\Ray=10^{10}$, $\Rey=5\,700$\\[3pt]
  \includegraphics[width=0.8\columnwidth]{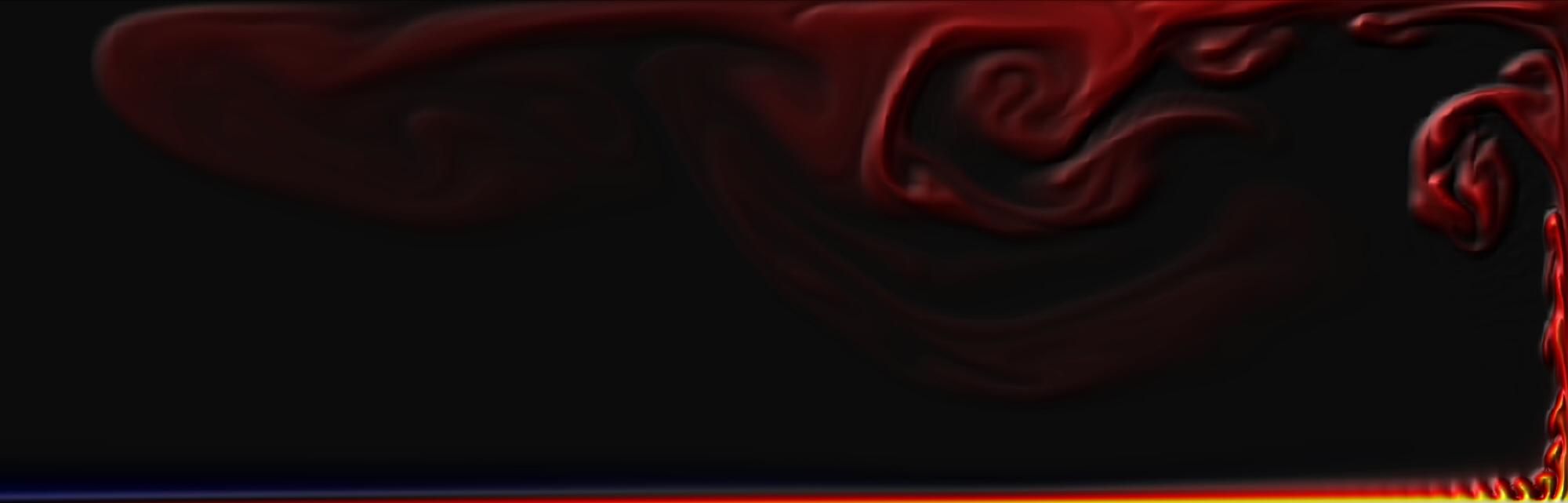} \\[6pt]
  (\textit{c}) Case~3: $\Ray=10^{10}$, $\Rey=7\,329$\\[3pt]
  \includegraphics[width=0.8\columnwidth]{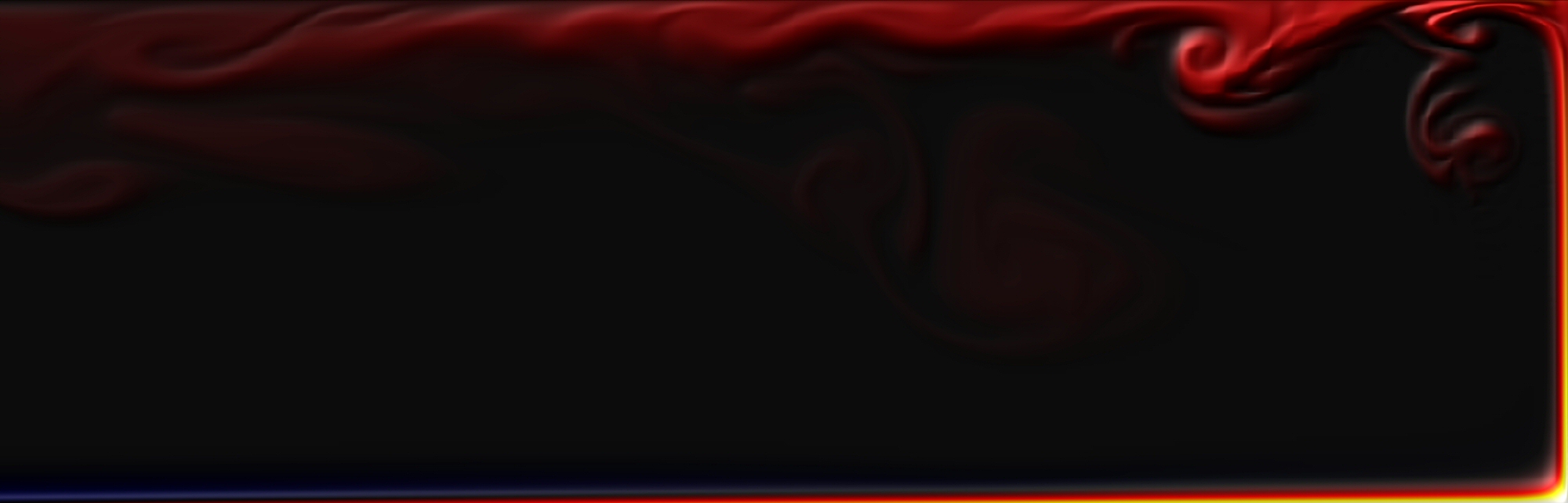}
\end{tabular}
 \caption{Instantaneous snapshots obtained at time $2.4\times 10^{-5}L^2/\kappa_T$ for the three cases displayed in the video. The right-hand (heated) half of the enclosure is shown, the forcing boundary is at the bottom of each frame, and black through red to yellow contours represent increasing temperature.}\label{fig:3cases}
\end{figure}

An explanation for the different behaviors observed in each of these sequences can be determined by considering the velocity boundary layers produced at the forcing boundary by each of these mechanisms. In a convection-dominated flow (\eg Case~1), the buoyancy destabilization invokes a boundary layer which features a tangential velocity profile with a positive gradient normal to the wall at the wall, a maximum velocity above the wall, before adopting a subsequent decrease at greater distances from the wall.  In other words, the buoyancy imbalance drives flow from left to right adjacent to the forcing boundary. In a flow dominated by mechanical forcing (\eg Case~3), the forcing boundary velocity profile is maximum at the wall, and decays further from the wall. In this case the wall movement leads and drives the flow through the action of viscous shear, carrying the temperature field in an almost passive fashion. Hence in this case, where both mechanisms are active, the flow will be dominated by natural convection when the boundary layer velocities invoked by buoyancy destabilization exceed the mechanical forcing boundary velocity.

A notable feature appearing in this video is an instability in the forcing boundary near the right-hand wall that manifests as a short-wavelength waviness that rapidly grows in amplitude and breaks into mushroom-plumes rising from the forcing boundary. These are invoked in cases where the horizontal convection mechanism has carried cooler fluid over hotter parts of the forcing boundary, creating a locally unstable flow with properties similar to classical Rayleigh--B\'{e}nard convection \cite{BodenschatzPeschAhlers-ARFM2000}, where convection cells are established to allow deeper hot fluid to displace cooler fluid at shallower depths.


\begin{thebibliography}{1}

\bibitem{BodenschatzPeschAhlers-ARFM2000}
E.~Bodenschatz, W.~Pesch, and G.~Ahlers.
\newblock Recent developments in {R}ayleigh-{B}\'{e}nard convection.
\newblock {\em Annu.\ Rev.\ Fluid Mech.}, 32:709--778, 2000.

\bibitem{HughGriff-2008-arfm}
G.~O. Hughes and R.~W. Griffiths.
\newblock Horizontal convection.
\newblock {\em Annu.\ Rev.\ Fluid Mech.}, 40:185--208, 2008.

\bibitem{MulGrifHugh-2004-jfm}
J.~C. Mullarney, R.~W. Griffiths, and G.~O. Hughes.
\newblock Convection driven by differential heating at a horizontal boundary.
\newblock {\em J.\,Fluid Mech.}, 516:181--209, 2004.

\bibitem{ShankarDeshpande-arfm2000-driven-cavity}
P.~N. Shankar and M.~D. Deshpande.
\newblock Fluid mechanics in the driven cavity.
\newblock {\em Annu.\ Rev.\ Fluid Mech.}, 32:93--136, 2000.

\bibitem{SheardKing2011amm-horiz-conv}
G.~J. Sheard and M.~P. King.
\newblock Horizontal convection: Effect of aspect ratio on {R}ayleigh-number
  scaling and stability.
\newblock {\em Appl.\ Math.\ Mod.}, 35(4):1647--1655, 2011.

\bibitem{TailleuxRouleau-TellusA-2010}
R.~Tailleux and L.~Rouleau.
\newblock The effect of mechanical stirring on horizontal convection.
\newblock {\em Tellus A}, 62(2), 2010.

\end{thebibliography}

\end{document}